\documentclass[twocolumn,superscriptaddress,showpacs,preprintnumbers,pra]{revtex4-1}
\usepackage{epsfig}
\usepackage{amssymb}
\usepackage{amsmath}
\usepackage{epstopdf}
\usepackage{graphicx} 
\usepackage{color} 
\usepackage[caption=false]{subfig}
\usepackage{bm}  
\begin{document}

\title{Bound states of a light atom and two heavy dipoles in two
  dimensions}

\author{D.~S. Rosa}
\address{Instituto de F\'\i sica Te\'orica, Universidade Estadual Paulista, Rua Dr. Bento Teobaldo Ferraz, 271 - Bloco II, 01140-070, S\~ao Paulo, SP, Brazil}
\author{F.~F. Bellotti}
\address{Department of Physics and Astronomy, Aarhus University, DK-8000 Aarhus C, Denmark}
\author{A.~S. Jensen}
\address{Department of Physics and Astronomy, Aarhus University, DK-8000 Aarhus C, Denmark}
\author{G. Krein}
\address{Instituto de F\'\i sica Te\'orica, Universidade Estadual Paulista, Rua Dr. Bento Teobaldo Ferraz, 271 - Bloco II, 01140-070, S\~ao Paulo, SP, Brazil}
\author{M.~T. Yamashita}
\address{Instituto de F\'\i sica Te\'orica, Universidade Estadual Paulista, Rua Dr. Bento Teobaldo Ferraz, 271 - Bloco II, 01140-070, S\~ao Paulo, SP, Brazil}

\begin{abstract}
We study a three-body system, formed by a light particle and two
identical heavy dipoles, in two dimensions in the Born-Oppenheimer
approximation.  We present the analytic light-particle wave function
resulting from an attractive zero-range potential between the light
and each of the heavy particles.  It expresses the large-distance
universal properties which must be reproduced by all realistic
short-range interactions.  We calculate the three-body spectrum for
zero heavy-heavy interaction as a function of light to heavy mass ratio.
We discuss the relatively small deviations from Coulomb estimates and
the degeneracies related to radial nodes and angular momentum quantum
numbers.  We include a repulsive dipole-dipole interaction and
investigate the three-body solutions as functions of strength and
dipole direction.  Avoided crossings occur between levels localized in
the emerging small and large-distance minima, respectively.  The
characteristic exchange of properties such as mean square radii are
calculated.  Simulation of quantum information transfer is suggested.
For large heavy-heavy particle repulsion all bound states have
disappeared into the continuum.  The corresponding critical strength
is inversely proportional to the square of the mass ratio, far from
the linear dependence from the Landau criterion.
\end{abstract}

\maketitle

%
%
%
%
%

\section{Introduction}

The experimental realization of ultracold atomic traps with the
further control of the interaction between the atoms using the
Feshbach resonance technique \cite{pethick,BloRMP2008,ChiRMP2010}
provided an unprecedented playground to study few-body correlations
\cite{NiePR2001,BraPR2006,BluRPP2012,FreFBS2011} that started long ago
uniquely as a theoretical study in the nuclear context
\cite{PhiNPA1968,CoePRC1970, TjoPL1975,EfiPL1970}. More specifically,
the tunability of the two-body interaction turned into a fact the
prediction made by Efimov that a system compounded by three identical
bosons presents an infinite number of bound states when the energy of
the two-boson subsystem tends to zero \cite{EfiPL1970}. This
interesting phenomenon occurs exclusively in three dimensions
\cite{NiePR2001} as a consequence of the collapse of the three-body
scale \cite{ThoPR1935}.

The first proposal about how to use magnetic fields in order to confine 
cold neutral atoms was made by Pritchard in 1983 \cite{PriPRL83} and its 
experimental observation was made two years later by Migdall \cite{MigPRL85}. 
Almost ten years after the first neutral atoms have been trapped, the question 
of dimensionality started to be explored in experiments \cite{SchAPB95}. 
Many aspects of the physics may change drastically when the system passes 
from three to two dimensions \cite{WerPRA2012,BelFBS2015,YamJPB2015}. 
Two pertinent examples can be cited here. The 
first one is the possibility to have an inherent attraction for zero orbital 
angular momentum that automatically appears from the kinetic energy. The 
consequence is that any infinitesimal attraction binds the system (in three 
dimensions the centrifugal barrier is always repulsive or null). The second example 
is related to the Efimov effect cited in the previous paragraph. In two 
dimensions the Thomas collapse is absent and no new scale should be added 
to the system at the universal regime: a two-body scale (the two-body 
energy, $E_2$, for example) defines completely the three-body observables, e.g. 
for three identical bosons there are only the ground and first excited 
states with energies satisfying, respectively, the relations $E_3^{(0)}=16.52E_2$ 
and $E_3^{(1)}=1.27E_2$ \cite{BruPRA1979}. The universal regime is accessed 
when the size of the system (a good quantity to describe it can be the two-body 
scattering length, $a$) is much greater than the range of the potential, 
$r_0$. In this regime, the observables do not depend on the details of 
the short-range potential.

The proportionality relations between $E_2$ and $E_3$ immediately shows that the 
infinite number of three-body bound states when $E_2=0$ is no longer possible in the 
bidimensional situation \cite{LimPRB1980,AdhPRA1988}. However, we can imagine 
a favorable design to generate 
more than two three-body bound states. It was derived long ago that for an 
asymmetric system $AAB$ formed by two identical bosons with mass $m_A$ 
and a different atom with mass $m_B$, it is possible to generate as many states 
as desired decreasing the ratio $ m_B/m_A$. The physical interpretation is 
that the effective potential becomes more attractive as the light atom 
can be easily exchanged by the heavy ones \cite{LimZPA1980}. The possibility to increase the 
attraction just by changing the mass ratio between the atoms creates an 
interesting situation: for any repulsion between the atoms of the heavy pair, 
we may always set a mass ratio in order to have a three-atom bound state.

There is still another interesting result associated with the
combination of both interactions: the sum of the effective and
repulsive potentials creates a barrier separating two minima each of them
supporting bound states with proper tuning.  The related tunneling can
be identified by the avoided crossing behavior of the energy
spectrum.  The impressive development of the experimental techniques
may then allow control over the localization of the wave function, and
in turn provide a laboratory to simulate transfer of quantum
information \cite{CirPRL1995,EckPRA2002,StrPRA2008,SodNJP2009,
  LadNature2010}.

The constant experimental advances of techniques involving ultracold
atomic traps in the last ten years produced a remarkable ability to
control the quantum properties of single atoms in an optical lattice
\cite{BakNature2008,WeiNature2011}.  More recently, the study of
dipoles became a hot topic in cold gases \cite{BarPR2008,
  LahRPP2009,KifPRL2013,HanPS2015} with a variety of experiments
producing traps of mixed-atoms and dipole molecules
\cite{NiScience2008,TakPRL2014,MolPRL2014,ParPRL2015,GuoPRL2016}.
These long-range interactions may be controlled by changing the
magnetic field or changing individually the magnetic moments of the
atoms. The dipole interaction can be made attractive or repulsive
depending on the orientation of the moments relative to their motion.
The short-range interactions may be suppressed by using the Feshbach
resonance to tune an infinity scattering length, and thereby producing a
pure dipole-dipole interacting two-body system.

The paper is organized as follows. In section \ref{formalism} we show
briefly the formalism used to calculate in two dimensions (2D) a
three-body system in the Born-Oppenheimer (BO) approximation formed by
two identical heavy dipoles and one atom. We fix the directions of the
dipoles in order to generate a repulsive or zero interaction between
them. The atom-dipole interaction is assumed to be of very short range
approximated by a zero-range interaction. We then derive the
analytical form of the light-atom wave function in the configuration
space which can be used as input to parametrize more realistic
calculations. In section \ref{resultsa} we discuss the solutions for
vanishing dipole interaction as function of the mass ratio.  In
section \ref{results} we investigate the solutions for finite dipole
interaction. We specially focus on strengths around degeneracy of
states localized in inner and outer minima, and finally we derive
critical dipole-dipole strengths for stability of the three-body
system.  A summary and conclusions are presented in section
\ref{conclusion}.

\section{Formalism}
\label{formalism}

\begin{figure}[htb!]
\includegraphics[width=8.5cm]{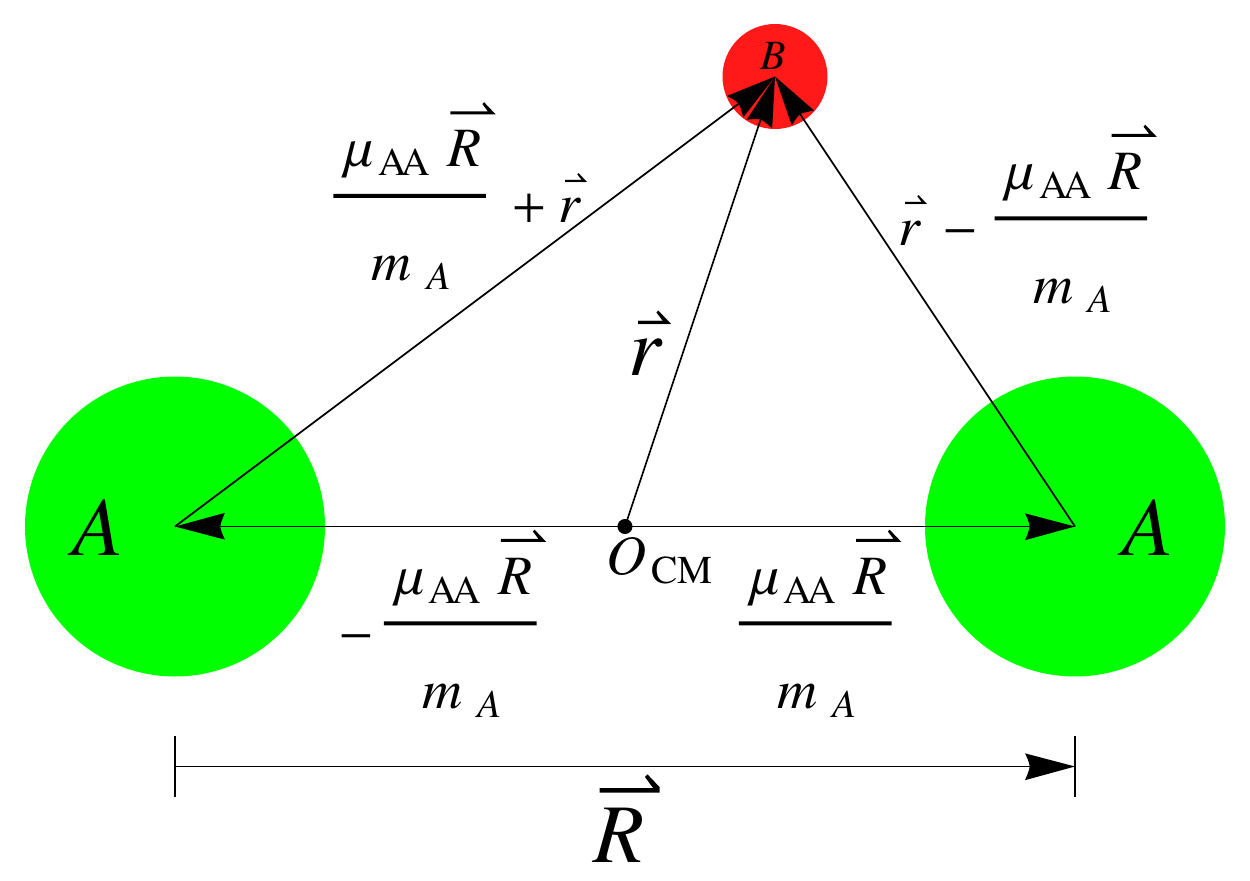}
\caption{ Three-body system formed by two identical dipoles with mass
$m_A$, $\mu_{AA}=m_A/2$, and an atom with mass $m_B$. We are 
considering the Born-Oppenheimer validity range where $m_B/
m_A \ll 1$. }
\label{figsystem}
\end{figure}

In this section we show the formalism used to solve the
three-body system formed by two heavy dipoles with masses $m_A$ and an
atom with mass $m_B$. We follow closely the theory from
Ref. \cite{BelJPB2013} repeating part of it in order to be
selfcontained.  We will consider the system depicted schematically in
Fig. \ref{figsystem}. The Hamiltonian is given by

\begin{eqnarray}
\label{hamiltonian}
H &=& -\frac{\hbar^{2}}{2\mu_{AA}}\nabla^{2}_{R}  -\frac{\hbar^{2}}{2\mu_{B,AA}}
\nabla^{2}_{r} + V_{B}(\vec{R}) \nonumber \\ 
&+&V_{A}(\vec{r}-\frac{\mu_{AA}}{m_{A}}\vec{R})
+ V_{A}(\vec{r}+\frac{\mu_{AA}}{m_{A}}\vec{R}),
\end{eqnarray}
where the reduced masses are given by $\mu_{AA}=m_A/2$ and
$\mu_{B,AA}=2m_A m_B/\left(2m_A+m_B\right)$.  
Here we are using an odd-man-out notation that is $V_A$ and 
$V_B$ denote, respectively, the $AB$ (atom/dipole) and $AA$ (dipole/dipole) 
two-body interactions. Note that we work with
relative coordinates after having removed the three-body
center-of-mass motion.

Let us consider here that $m_A\gg m_B$. Due to this severe mass asymmetry we may 
consider the dipoles stationary when we solve the problem of the atom $B$. The 
total wave function may be written as a product of the wave functions of the fast 
atom, $\psi$, and slow dipoles, $\phi$, as 
$\Psi(\vec{r},\vec{R})=\psi(\vec{r},\vec{R})\phi(\vec{R})$, where the separation 
vector of the dipoles, $\vec{R}$, enters in $\psi$ only as a parameter. 
It is worth mentioning here that the total wave function 
is symmetric under the interchange of the heavy dipoles. As the interaction 
is not dependent on the spin, the formalism developed here may be suitable 
to describe a three-body system formed by bosonic or antiparallel fermionic 
dipoles.

The action of the Laplace operator, $\nabla_{R} ^{2}$, on $\Psi$ in
the Schroedinger equation is to lowest order approximated by
$\nabla_{R} ^{2}\phi(\vec{R})\psi(\vec{r},\vec{R}) \approx \psi(\vec{r},\vec{R})
\nabla_{R} ^{2}\phi(\vec{R})$.  Two other terms, $ \phi(\vec{R})
\nabla_{R} ^{2}\psi(\vec{r},\vec{R})$ and $2 \nabla_{R}
\psi(\vec{r},\vec{R}) \cdot \nabla_{R} \phi(\vec{R})$, also appear but
usually they are much smaller and we neglect them in the present
paper.  We are then able to write independent equations for the heavy and
light subsystems as
\begin{eqnarray}
&&\left[ -\frac{\hbar^{2}}{2\mu_{B,AA}}\nabla^{2}_{r} + V_{A}(\vec{r}-\frac{\mu_{AA}}{m_{A}}\vec{R}) \right. \nonumber \\ 
&&+ \left. V_{A}(\vec{r}+\frac{\mu_{AA}}{m_{A}}\vec{R})\right]\psi(\vec{r},\vec{R}) = \epsilon(R)\psi(\vec{r},\vec{R}), 
\label{light}\\
&&\left[-\frac{\hbar^{2}}{2\mu_{AA}}\nabla^{2}_{R}  + V_{B}(\vec{R}) + \epsilon(R) \right]\phi(\vec{R}) = 
 E_3 \phi(\vec{R}).
\label{heavy}
\end{eqnarray} 
Note that the eigenvalue of Eq. (\ref{light}), $\epsilon(R)$, depends
on the relative position of the heavy dipoles, $R$, and enters as an
effective potential in Eq. (\ref{heavy}). The eigenvalue of this heavy
dipoles equation returns the three-body binding energy, $E_3$.
\vspace{1cm}

\subsection{Light-particle equation}

The form of $\epsilon(R)$ was firstly derived in Refs. \cite{LimZPA1980} for two different Yamaguchi 
form factors. Further, in Ref. \cite{BelJPB2013}, this effective potential was calculated for a 
zero-range potential. It results from a solution of a transcendental equation given by

\begin{equation}
\ln \left[\frac{|\epsilon(R)|}{|E_{2}|}\right] = 2 K_{0}\left(\sqrt{\frac{2\mu_{B,AA} |\epsilon(R)|}{\hbar^{2}}}R \right),
\label{epsilon}
\end{equation}
where $K_{0}$ is the zero order modified Bessel function of the second kind. Solving this equation for 
$\epsilon(R)$ we see that the attraction increases when the mass ratio $m_B/m_A$ is decreased, as the 
light atom, which generates the effective attraction, can be more easily exchanged between the dipoles. 
This potential as well as the wave function in momentum space, $\psi(\vec{p},\vec{R})$ 
($\vec{p}$ is the momentum canonically conjugate to $\vec{r}$), appeared in Ref. \cite{BelJPB2013}. 
The Fourier transform of $\psi(\vec{p},\vec{R})$ can be analytically calculated such that in 
coordinate space the light-atom wave function is given by:

\begin{eqnarray}
\psi(\vec{r},\vec{R}) &=& - \frac{\mu_{B,AA}}{\hbar^{2}\pi}  
\left[ K_{0}\left( \sqrt{\frac{2\mu_{B,AA}|\epsilon(R)|} {\hbar^{2}} } 
\rvert \vec{r}+\frac{\mu_{AA}}{m_{A}}\vec{R} \rvert \right) \right. \nonumber \\
&+& \left. K_{0}\left(\sqrt{ \frac{2\mu_{B,AA}|\epsilon(R)|} {\hbar^{2}}} \rvert 
\vec{r}-\frac{\mu_{AA}}{m_{A}}\vec{R} \rvert  \right)\right]. 
\label{lightwavefunction}
\end{eqnarray}

The atom-dipole binding energy, $E_2$, for a renormalized zero-range interaction enters as an input 
in the effective potential, $\epsilon(R)$,
see Eq. (\ref{epsilon}). The form of the wave function given in 
Eq. (\ref{lightwavefunction}) can be used to parametrize the 
light-heavy large-distance tail of the wave 
function of any potential 
satisfying the Born-Oppenheimer validity condition.

\subsection{Heavy particles equation}

In the next section we solve Eq. (\ref{heavy}) numerically. Firstly, 
we change variables based on the small distance behavior of 
the effective potential. Note that for small distances, the effective potential 
is given by \cite{BelJPB2013}
\begin{equation}
\epsilon(R) \rightarrow - \frac{2e^{-\gamma}|E_{2}|}
{\sqrt{\frac{2\mu_{B,AA}|E_{2}|}{\hbar^{2}}}}\frac{1}{R}
\ \ {\rm for}\ \ \sqrt{\frac{2\mu_{B,AA}|E_{2}|}{\hbar^{2}}} 
R \leq 1.15,\label{short}
\end{equation}
where $\gamma=0.5772$ is Euler's constant. This Coulomb-like short-range behavior 
forces the system to act like a hydrogen atom with an effective squared charge given by 
\begin{equation}
\mathcal{Q}_{\rm{eff}}^{2} =2\sqrt{\frac{\hbar^{2}}{2\mu_{B,AA}|E_{2}|}}
e^{-\gamma} |E_{2}|,
\label{qeff}
\end{equation}
and the corresponding modified Bohr radius
\begin{eqnarray}
a_{0}&=&\frac{\hbar^{2}}{\mu_{AA}\mathcal{Q}_{\rm{eff}}^{2} }.
\label{aff} 
\end{eqnarray}
We can then rewrite Eq. (\ref{heavy}), after  making the change of variable 
from $R$ to $x$, that is $R = \left(a_0/2\right) x$, as
\begin{eqnarray}
\nonumber
&& \left[ \frac{\partial^{2}}{\partial x^{2}} - \frac{(l^{2}-1/4)}{x^{2}}  \right] 
\chi(x) +\frac{a_0}{2\mathcal{Q}_{\rm{eff}}^{2}}\left[ \epsilon(x)-V_{B}(x) 
\right]\chi(x) \\ \label{heavyfinal}
&&= -\frac{a_0}{2\mathcal{Q}_{\rm{eff}}^{2}}E_3 \chi(x),
\end{eqnarray}
where $\chi(x) = \sqrt{R} \phi(R)$, $V_B$ is the heavy-heavy particle
potential, and the non-negative integer, $l$, is the angular momentum
quantum number in two dimensions arising from the kinetic energy
operator in spherical coordinates.  It is worth noting here that
Eq. (\ref{heavyfinal}) shows an important difference when considering
three-dimensional (3D) or 2D systems, namely, for $l=0$ we have here an attractive centrifugal
barrier, while in 3D the barrier is always repulsive or null.

\section{The three-body Born-Oppenheimer structure}
\label{resultsa}

We shall first discuss the Born-Oppenheimer wave function for a fixed
distance between the two heavy particles. In the next subsection we
present the basic three-body spectra for vanishing heavy-heavy
particle potential.

\subsection{Light particle wave function }

The short-range potential between light and heavy particles is
approximated by a zero-range interaction. This allows
an analytic solution with the wave function in coordinate space
explicitly given in Eq. (\ref{lightwavefunction}).  Rewriting the
arguments in the Bessel function and using $\mu_{AA} = m_A/2$, we get
\begin{eqnarray} \label{wavelight}
 \psi(\vec{r},\vec{R}) &=& - \frac{\mu_{B,AA}}{\hbar^{2}\pi}  
  \left[K_{0}(\zeta_{+}) + K_{0}(\zeta_{-} )\right] \;,
\end{eqnarray}
where the wave function is seen to be a function of two coordinates
and one scale parameter $b$, that is
\begin{eqnarray} \label{escale}
\zeta_{\pm}  &=& \sqrt{b\left(\frac{r^2}{R^2} + \frac{1}{4} 
 \pm \frac{r}{R} \cos(\theta_{rR}) \right) } \;, \\
 b  &=&  \frac{2\mu_{B,AA} R^2 |\epsilon(R)|} {\hbar^{2}}.   
\label{besarg}
\end{eqnarray}
Thus, the dimensionless combination in $b$ of energy, reduced mass
and heavy-heavy distance determines this wave function completely
together with the relative size, $r/R$, and the direction,
$\cos(\theta_{rR})$, between the two relative coordinates, $\vec{r}$
and $\vec{R}$, see Fig. \ref{figsystem}.

\begin{figure}[htb!]
\includegraphics[width=8.6cm]{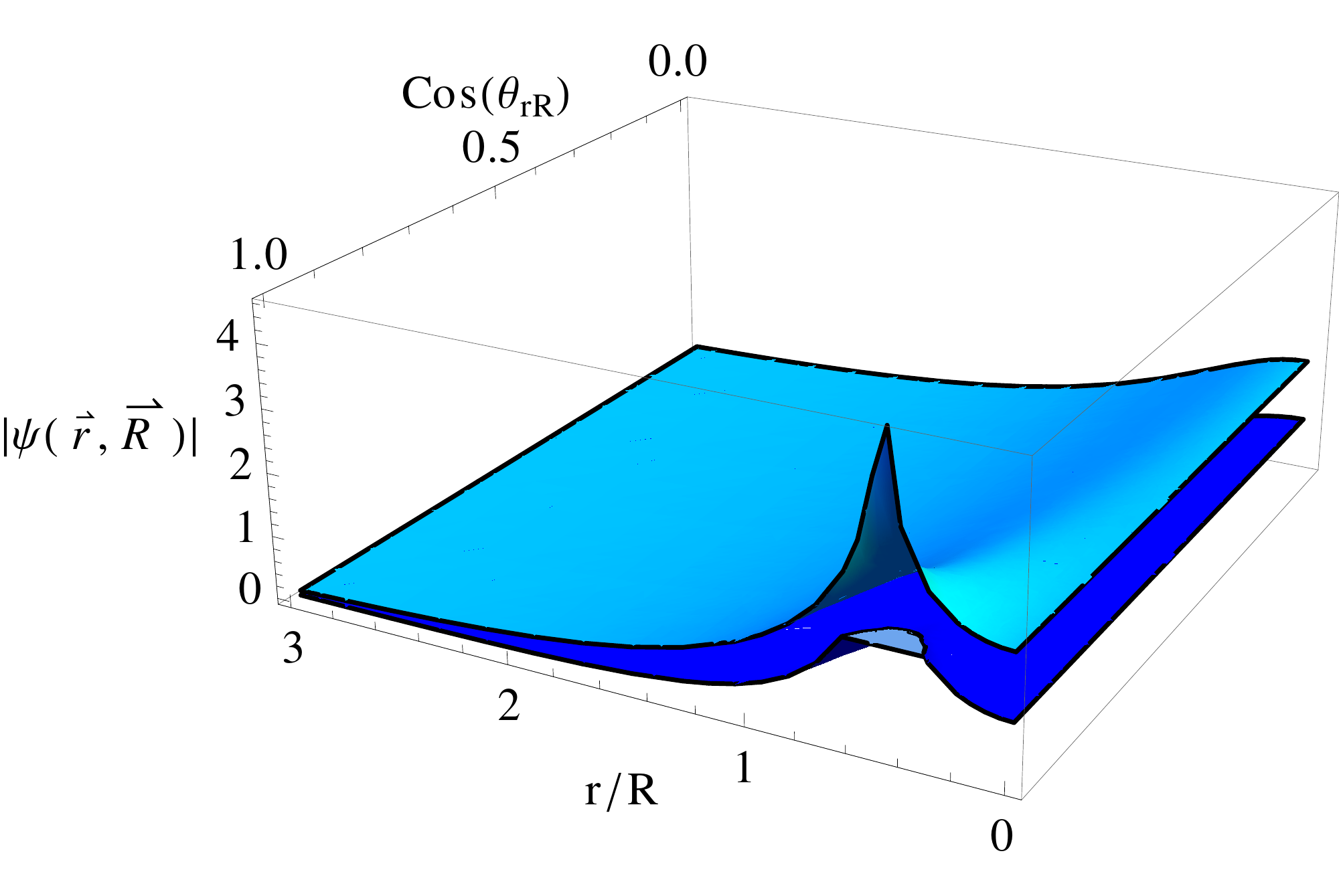}
\vspace{0.1cm}
\caption{The wave function in Eq. (\ref{wavelight}) as a function of the
  relative size, $r/R$, and the direction, $\cos(\theta_{rR})$,
  between the two relative coordinates, $\vec{r}$ and $\vec{R}$, see
  Fig. \ref{figsystem}. The upper and lower surfaces are, respectively, 
  results for $b = 1$ and $4$. The constant in front of Eq. (\ref{wavelight}) 
is fixed as $\mu_{B,AA}/(\hbar^{2}\pi)=1$.}
\label{psiwave}
\end{figure}

The wave function $\psi$ is symmetric around $\cos(\theta_{rR}) =
0$.  The Bessel function has a logarithmic divergence when its
argument aproaches zero, which in the present case occurs when $R = 2
r$ and $\cos(\theta_{rR}) = \pm 1$. This is precisely when the light
particle is on top of one of the heavy particles.  For the employed
short-range attraction this is not a surprise as the probability
is largest in such a situation, as seen in Fig. \ref{psiwave}.  The two surfaces for
different $b$ shows how the peak increases around the heavy
particles as the two-body binding energy or heavy-heavy particle
distance increases.  The light particle becomes in both cases
increasingly localized around one of the heavy particles, the effect of 
increasing $b$ is to make the wave function flatter.

The properties seen in Eq. (\ref{wavelight}) and illustrated in
Fig. \ref{psiwave} reflect the genuine universal behavior for any
short-range interaction in the Born-Oppenheimer approximation,
provided the distances between light and heavy particles are much
larger than the range of the respective potentials.  This analytic form of the
wave function can then be used as a boundary condition or to parametrize
the tail of a general wave function.

\subsection{Heavy particle motion }

We now turn to the three-body bound-state properties obtained from
Eq. (\ref{heavyfinal}). Initially, we consider noninteracting two heavy
particles, $V_B=0$.  The solutions are completely determined from the
two-body energy, $E_2$, as functions of mass ratio, $m_B/m_A$, and
angular momentum $l$.  From Eq. (\ref{short}),
one can see that the effective potential for small distances behaves like a
Coulomb potential.  It can be rewritten in the new variables as
\begin{equation}
 \epsilon(x) \rightarrow - |E_{2}|\left(\frac{\mu_{AA}}{\mu_{B,AA}}\right)
 \frac{4e^{-2\gamma}}{x}\  \ {\rm for} \ \left(\frac{\mu_{B,AA}}{\mu_{AA}}\right)x \leq 1.29.
\label{shortnew}
\end{equation}
The corresponding energy solutions, are given analytically by \cite{SaPRA2016}
\begin{equation}
 \frac{|E_{3}^{(C)}|}{|E_{2}|} = -\frac{1}{2(n+l-\frac{1}{2})^{2}}
 \left(\frac{\mu_{AA}}{\mu_{B,AA}}\right)2e^{-2\gamma}, 
 \label{coulombexact}
\end{equation}
where we inserted the effective charge from Eq. (\ref{qeff}).  The
quantum numbers, $n$ and $l$, are both non-negative integers.  The
approximation in Eq. (\ref{coulombexact}) contains the usual Coulomb
degeneracy, where only the non-negative integer combinations, $n+l$,
determine the energies.  The mass dependence as well as the two-body
interaction strength, $E_2$, are very trivial in this approximation.

\begin{figure}
{  \includegraphics[width=8.6cm]{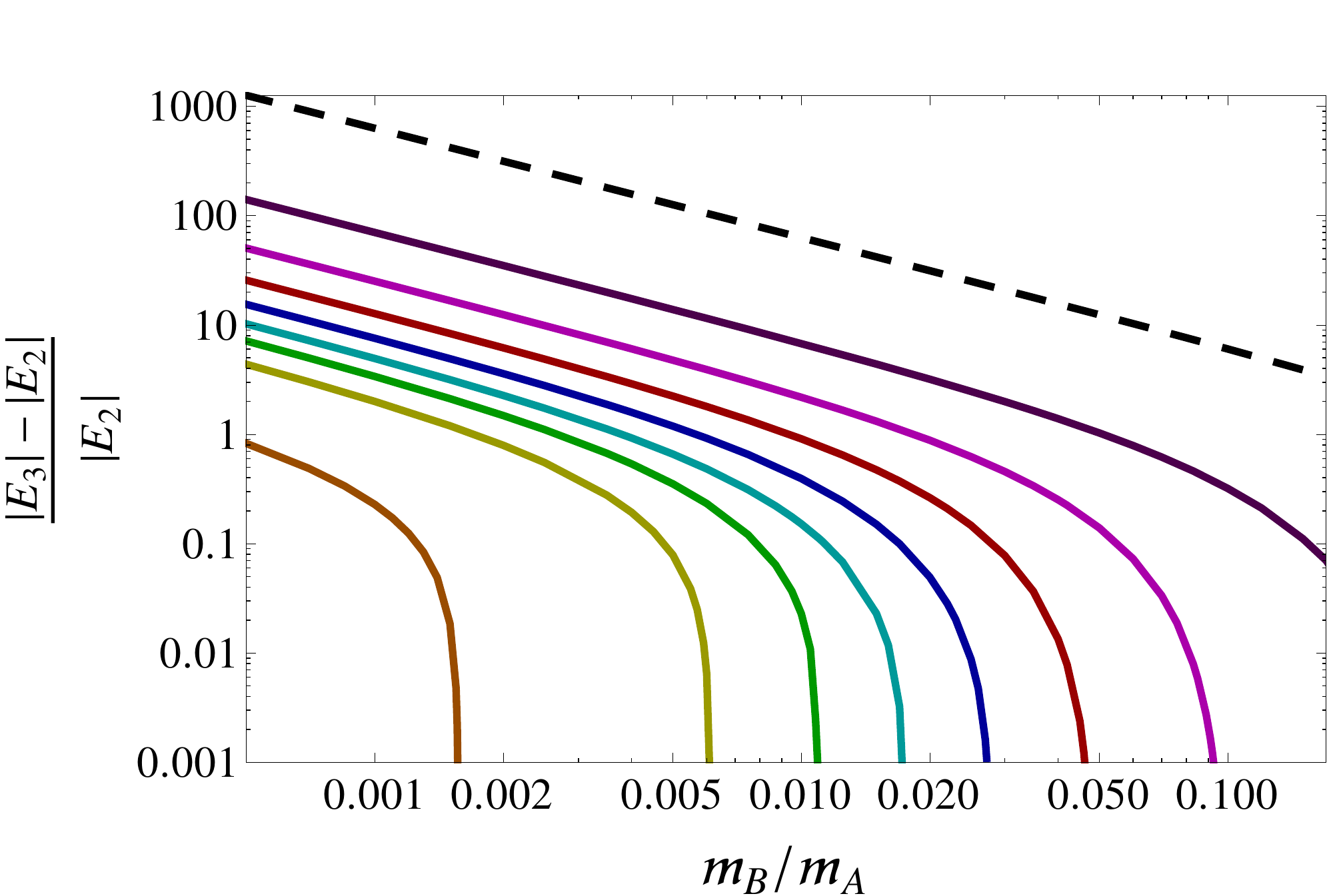}}
\hfill
\caption{Three-body energy, $E_3$, with respect to the two-body
  threshold, $E_2$, as a function of the mass ratio of the atom
  ($m_B$) and the heavy dipoles ($m_A$) for orbital angular
  momenta ($l=0$).  The dipole-dipole interaction, $V_B$, is zero. The
  dashed line is the ground state and the lines below it are,
  in sequence, results for the following excited states.  The first
  and second of these excited $l=0$ states coincide within the line
  thickness with the ground states for $l=1$ as and $l=2$,
  respectively.}
\label{fignointeraction}
\end{figure}

In Figure \ref{fignointeraction} we show the lowest three-body
energies calculated numerically from Eq. (\ref{heavyfinal}) for
$V_B=0$ as function of the mass ratio $m_B/m_A$.  Each energy
decreases with increasing $m_B/m_A$ as predicted by
Eq. (\ref{coulombexact}), and eventually reaches the
continuum at individual threshold values.  The number of bound states
is correspondingly reduced with $m_B/m_A$ as also known in general from
\cite{BelJPB2013}.  The threshold behavior of both energy and
structure is discussed in some details in \cite{SaPRA2016}.

The results for higher values of $l$ are almost indistinguishable
within the thickness of the curves, except that the largest binding energy
moves towards the next excited state for each unit of $l$.  This is precisely 
the Coulomb degeneracy in two spatial dimensions as contained in
Eq. (\ref{coulombexact}).  The ground state has $(n,l)=(0,0)$, the
first doubly degenerate excited energy has $(n,l)=(1,0),(0,1)$, and
the second $(n,l)=(2,0),(1,1),(0,2)$,  etc.  Thus,  the number of bound
states also decreases with $l$.  

This degeneracy is quantified in Table \ref{tabcoulombl0} where 
our numerical results are compared with the analytic ones given
by Eq. (\ref{coulombexact}) for $l=0$ and $m_B/m_A=0.02$. As mentioned above
the same numbers appear in comparison for larger $l$-values.  We
notice that the ratio of the numerically computed binding and 
the Coulomb values are $1.02$, $1.19$, $1.48$, and $1.96$ for
increasing excitation energy, respectively.  The principal reason is
that Eq. (\ref{coulombexact}) only is the lowest order in an expansion
for small $x$, see \cite{BelJPB2013}.  The next two terms contribute
although varying only weakly with $x$, that is, the first is simply a
constant and the second correction term is logarithmic in $x$.
They both provide larger binding energy.

\begin{table}[h]
\caption{Results for several states of the three-body energy in units
  of $E_2$ considering the BO approximation,
  $|E_{3}|/|E_{2}|$, and the exact hydrogen-atom calculation,
  $|E_{3}^{(C)}|/|E_{2}|$, for $l=0$. The mass ratio is fixed
  to $m_B/m_A=0.02$.}
\begin{tabular}{c c c c}
\hline\hline
{\rm State}  &\ \ \ \  $|E_{3}|/|E_{2}|$ \ \ \ &\  $|E_{3}^{(C)}|/|E_{2}|$  \ &\ \ \ $(|E_{3}|-|E_{3}^{(C)}|)/|E_{2}|$ \\ [0.9ex]  
\hline  
Ground       &32.42 & 31.83 &  0.59   \\
First        &4.19  & 3.53  & 0.66  \\
Second       &1.88  & 1.27  &  0.61 \\
Third        &1.26  & 0.64  &  0.62 \\[1ex]
\hline\hline
\end{tabular}
\label{tabcoulombl0}  
\end{table}

We emphasize that this discussion would be less and less relevant
for excited states approaching the thresholds for binding. Then the
deviations from the Coulomb estimate would arise from the extension of
the states into a region with strongly deviating Coulomb potential.
This happens by increasing excitation energy, increasing mass ratio,
and increasing angular momentum.

\section{Heavy particle motion with additional interaction}
\label{results}

In this section we investigate the three-body solutions with an
explicit heavy-heavy potential included on top of the
Born-Oppenheimer potential $\epsilon(R)$ arising from the zero-range light-heavy
potential.  This heavy-heavy potential could in principle be of any
form and strength; for example, extremely short- or long-range
potentials.  We shall here discuss only the realistic dipole-dipole
potential of intermediate range.  First we present the general
analytic and numerical properties and in the last subsection we
discuss the necessary critical strength to reach instability.

\subsection{Dipole-dipole interaction  }

Generically, for two dipoles, $m_1$ and $m_2$, connected by a vector
$\vec{R}$, the dipole interaction is given by \cite{HanPS2015}

\begin{eqnarray} \label{dippot}
V_B(\vec{R})=C\left[\frac{\vec{m}_1\cdot\vec{m}_2}{R^3}-
\frac{3(\vec{m}_1\cdot\vec{R})(\vec{m}_2\cdot\vec{R})}{R^5}\right],
\end{eqnarray}
where $C=\mu_0/4\pi$ and $C=1/4\pi\epsilon_0$ for
magnetic and electric dipoles, respectively (note that in Gaussian 
units the first and second constants should be replaced, respectively, by 
$C=1/c^2$ and $C=1$, where $c$ is the speed of light). Here $\mu_0$ is the
vacuum permeability and $\epsilon_0$ the vacuum permittivity.  For two
identical dipoles in the $xy$-plane with dipole moments $\vec{D}$
forming an angle $\theta$ with the $z$-axis and an azimuthal angle
$\phi$ we rewrite Eq. (\ref{dippot}) into
\begin{equation}
V_B(R,\theta,\phi)=\frac{CD^2}{R^3}(1-3\sin^2\theta\cos^2\phi),
\label{dipole}
\end{equation}
where $\vec{R}$ is along the $x$-direction.  This interaction,
is repulsive for two dipoles in a plane when $\vec{D}$ is
perpendicular to this plane. However, as the angle $\theta$ increases
the repulsion decreases until it vanishes and eventuallly becomes
attractive when $\vec{D}$ is a vector in the $xy$-plane.  The dipole
direction would in practice be determined by an external polarizing
field.  We shall in the following maintain arbitrary but fixed dipole
direction.

Changing the heavy-heavy relative coordinate to $x$ by $R=(a_0/2)x$, we
have
\begin{equation}
V_B(x,\theta,\phi)=\frac{8CD^2}{a_0^3x^3}(1-3\sin^2\theta\cos^2\phi),
\label{repulsion}
\end{equation}
which has a cubic divergence at $x=0$. This is both impractical as
well as unphysical since sufficiently small $x$ means distances where
chemical degrees of freedom becomes unavoidable \cite{AvanJPA2003}.  We therefore
regularize in the smoothest possible way by modifying for distances
$x$ smaller than a constant $x_0$, which can be thought of as
determined by the van der Waals \cite{BraPR2006} or dipole length \cite{BohNJP2009}.  
The full potential in Eq. (\ref{heavyfinal}) is then correspondingly given by
\begin{eqnarray}
 &&U(x)=\frac{a_0}{2\mathcal{Q}_{\rm{eff}}^2}\epsilon(x)+
 \lambda x_0^3\frac{1}{x^3+x_0^3} \; , \label{totalpot} \\
 &&\lambda \equiv 4e^{-\gamma}CD^{2}\sqrt{\frac{2\mu_{AA}^{4}|E_{2}|}{\mu _{B,AA}\hbar^{6}}}
 (1-3\sin^2\theta\cos^2\phi),
\label{ux}
\end{eqnarray}
where the strength, $\lambda$, now is dimensionless.

The Born-Oppenheimer part of the potential is attractive in all points
of space.  It is Coulomb-like at small distances and essentially
exponentially vanishing at large distances.  The Coulomb strength and
the exponential cut-off radius are both increasing with decreasing
mass ratio $m_B/m_A$.  Thus, both small and large distances allow more and more
bound states as that mass ratio decreases. 

Adding an overall repulsive potential without any divergence still
leaves the Coulomb behavior at small distance but not necessarily
allowing any bound states.  A relatively short-range repulsion leaves
on the other hand the large-distance attraction much less affected.
However, a sufficient repulsive strength on, for example, the dipole
potential must eventually remove all attraction and thereby all bound
states.  Intermediate strengths then allow two regions where bound states may exist 
at both small and large distances separated by a barrier, see Figure \ref{potential}.

\subsection{Three-body properties }

\begin{figure}[htb!]
\hspace{-0.5cm}\includegraphics[width=9.1cm]{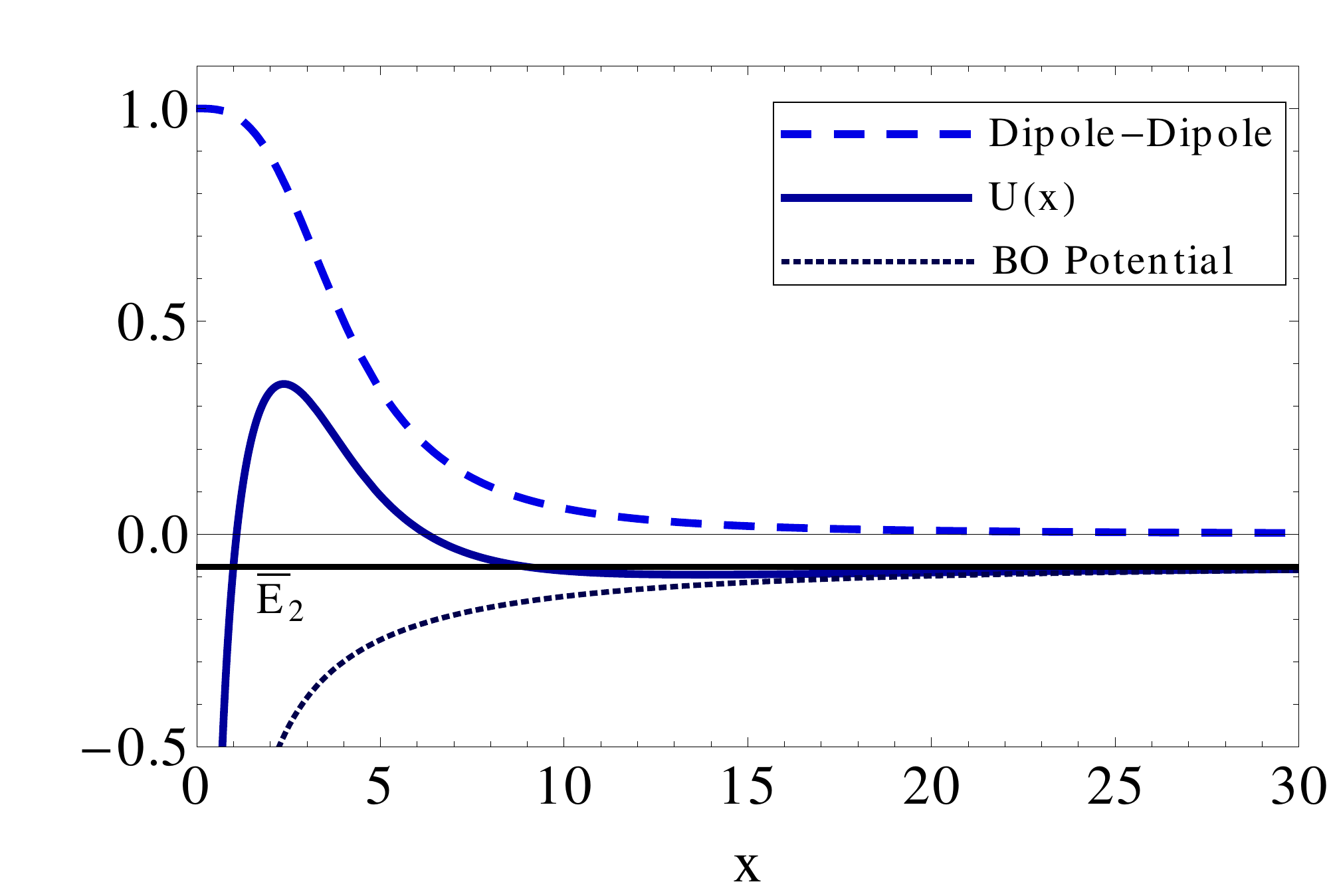}
\caption{The Born-Oppenheimer potential (dashed) and dipole-dipole
  potential (dotted) for $\lambda =1$, and total potentials (solid). 
  The horizontal lines are the two-body threshold ${\bar E}_2\equiv \left(a_0/2\mathcal{Q}_{\rm{eff}}^{2}\right)E_2$.  All potentials are plotted as
  functions of $x$, and calculated for $x_0=4.0$ and mass ratio $m_B/m_A =0.05$.}
\label{potential}
\end{figure}

\begin{figure}[h!]
\centering
\subfloat[]{  \includegraphics[width=8.5cm]{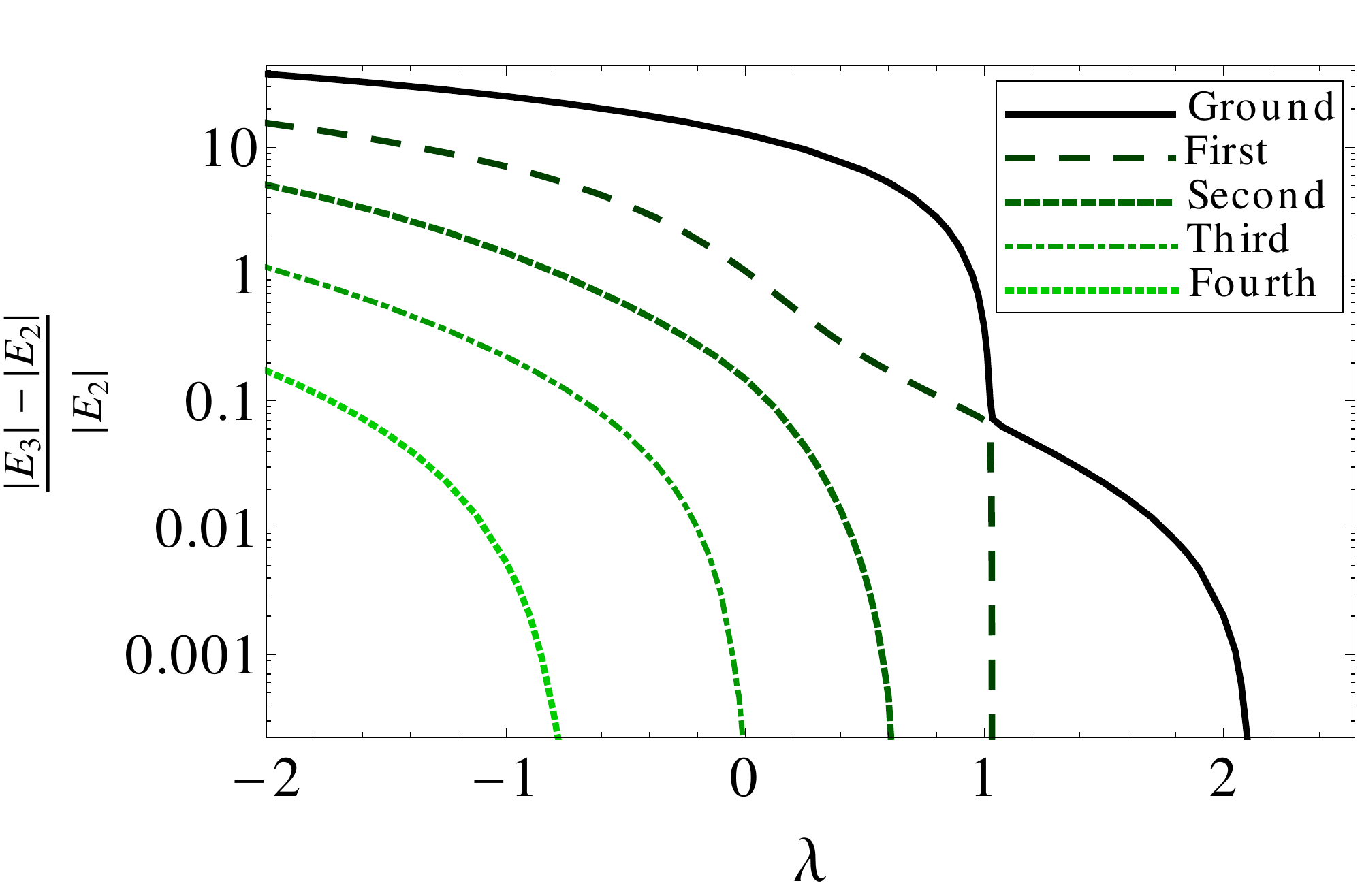}}
\\
\subfloat[]{  \includegraphics[width=8.5cm]{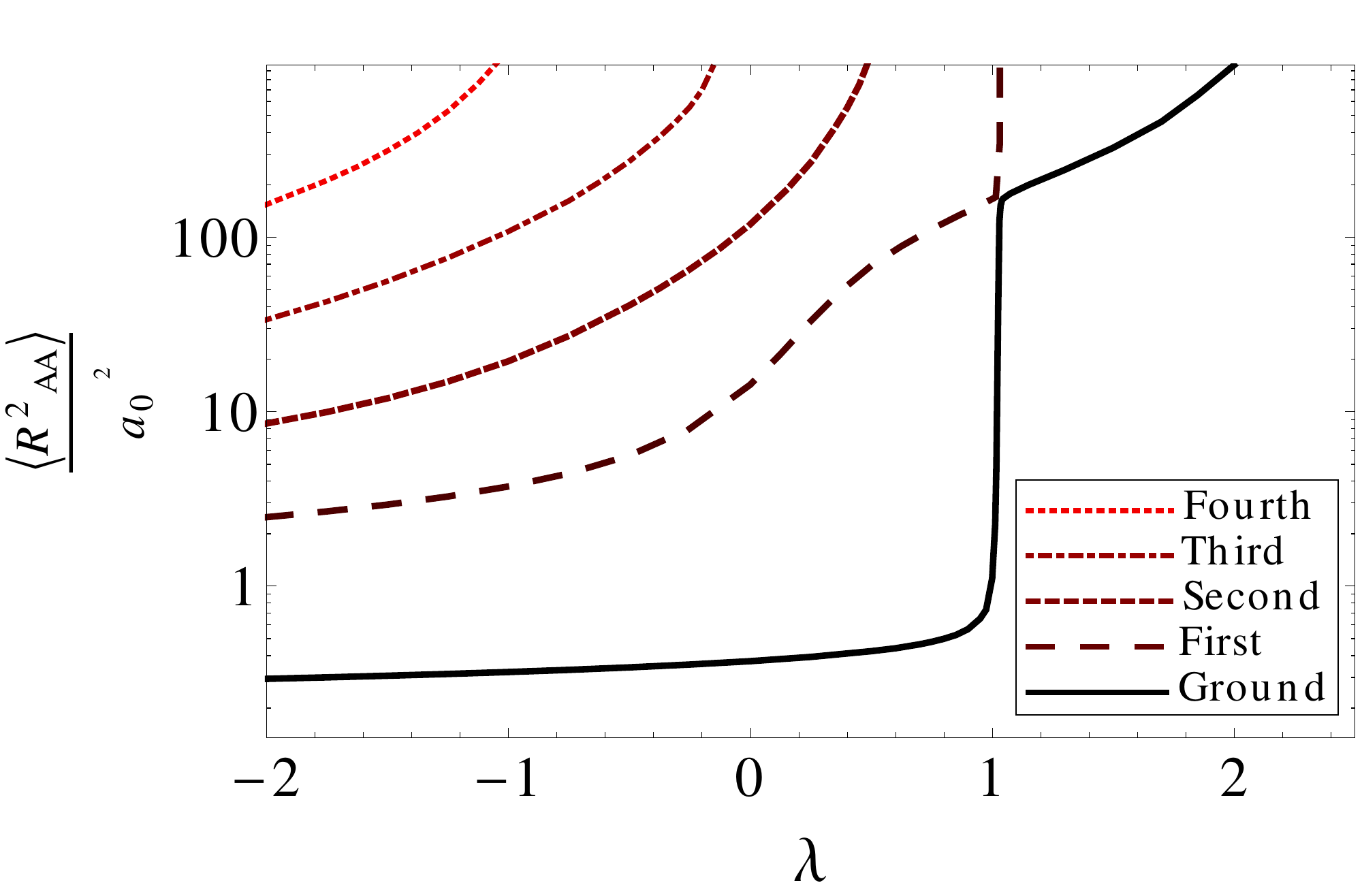}}
\caption{(a) Three-body energy spectrum, $E_3$, with respect to the
  two-body threshold, $E_2$, and (b) mean-square radii of the
  heavy-heavy distance for the respective states.  Both energies and
  sizes as functions of the strength, $\lambda$, of the dipole
  potential for a fixed $x_{0}=4.0$ and mass ratio $m_B / m_A = 0.05$.}
\label{figdip}
\end{figure}

The total three-body potential in Eq. (\ref{totalpot}) is shown in
Figure \ref{potential} for one dipole strength $\lambda$.  The eigenvalues and
mean-square radii are shown in Figure \ref{figdip} as functions of the
strength.  We first note the overall trend of decreasing energies and
increasing radii with increasing repulsion.  All higher-lying states,
except ground and first excited, vary smoothly with the dipole
strength.  They are located in the outer minimum until they are pushed
into the continuum by the repulsion.

In sharp contrast, the ground and first excited states exhibit around
$\lambda\approx 1$ the typical behavior of states avoiding to cross
each other. These two states are for small positive $\lambda$,
respectively located in the inner and outer ``minima'' as reflected by their
radii in Figure \ref{figdip}.  As $\lambda$ increases towards $1$, the
ground state is affected more than the first excited state and,
eventually, overtakes the first excited state.

At the degeneracy strength any linear combination of the two states
are valid solutions.  After the crossing they exchange structure
properties, but very quickly afterwards the highest-lying abruptly
disappear into the continuum.  There is not sufficient space in the
inner minimum to hold a bound state, whereas the lowest-lying state
remains relatively unaffected in the outer minimum.

\begin{figure*}
\hspace{-0.8cm}\subfloat[$\lambda  = 0.95$]{  \includegraphics[width=9cm]{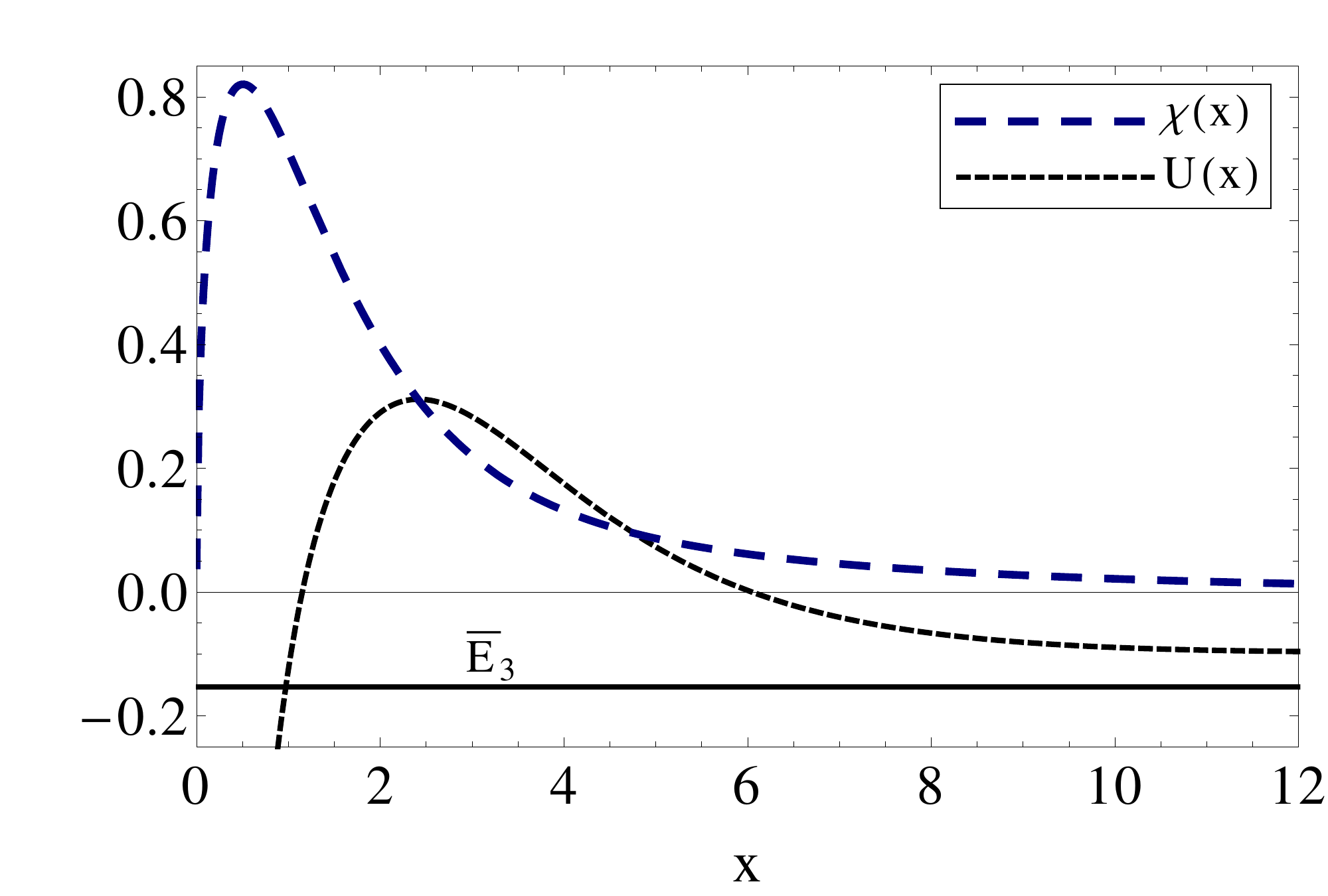}}\hspace{-0.25cm}
\subfloat[$\lambda  = 1.0$]{  \includegraphics[width=9cm]{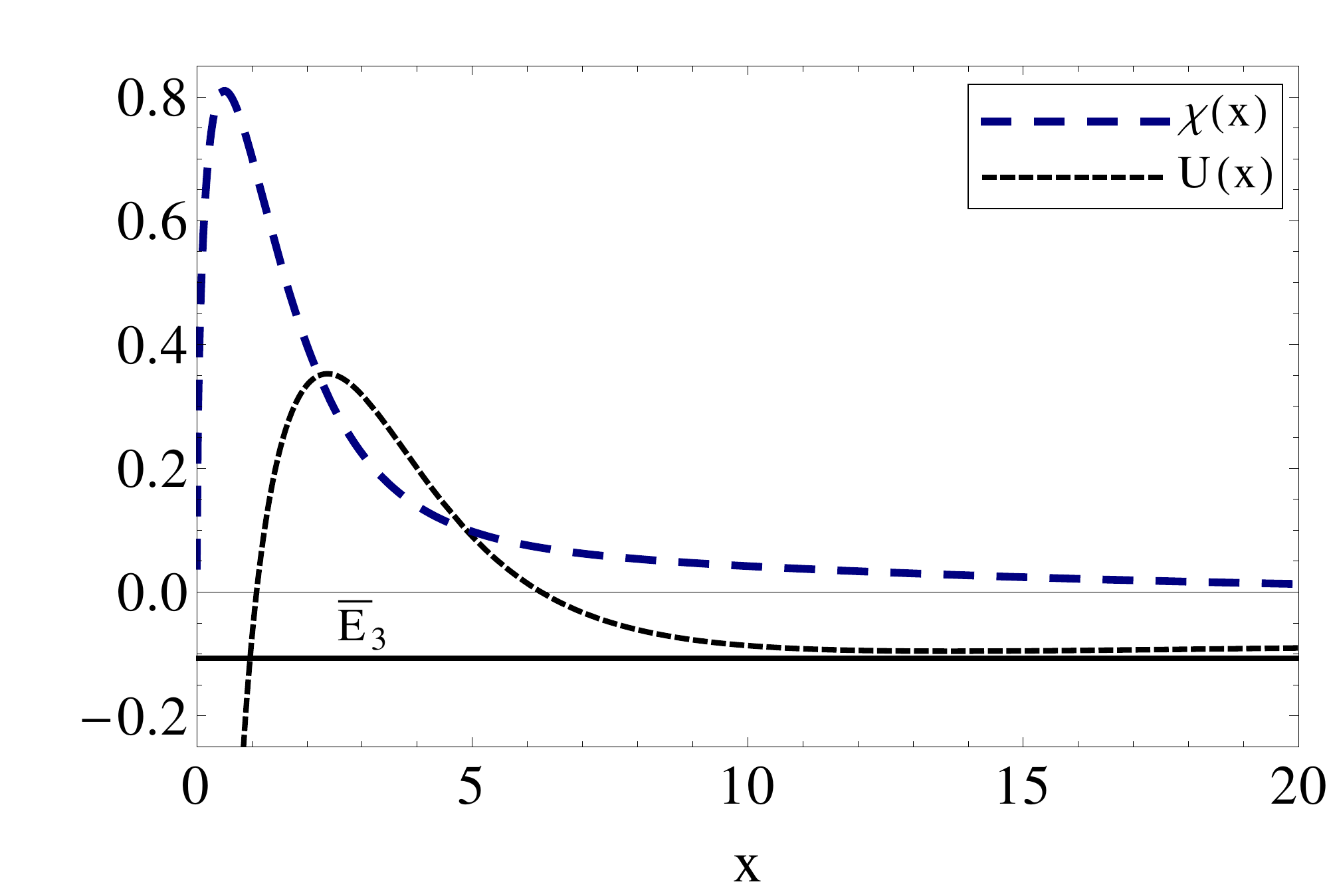}}
\\
\vspace{0.3cm}
\hspace{-0.8cm}\subfloat[$\lambda  = 1.012$]{  \includegraphics[width=9cm]{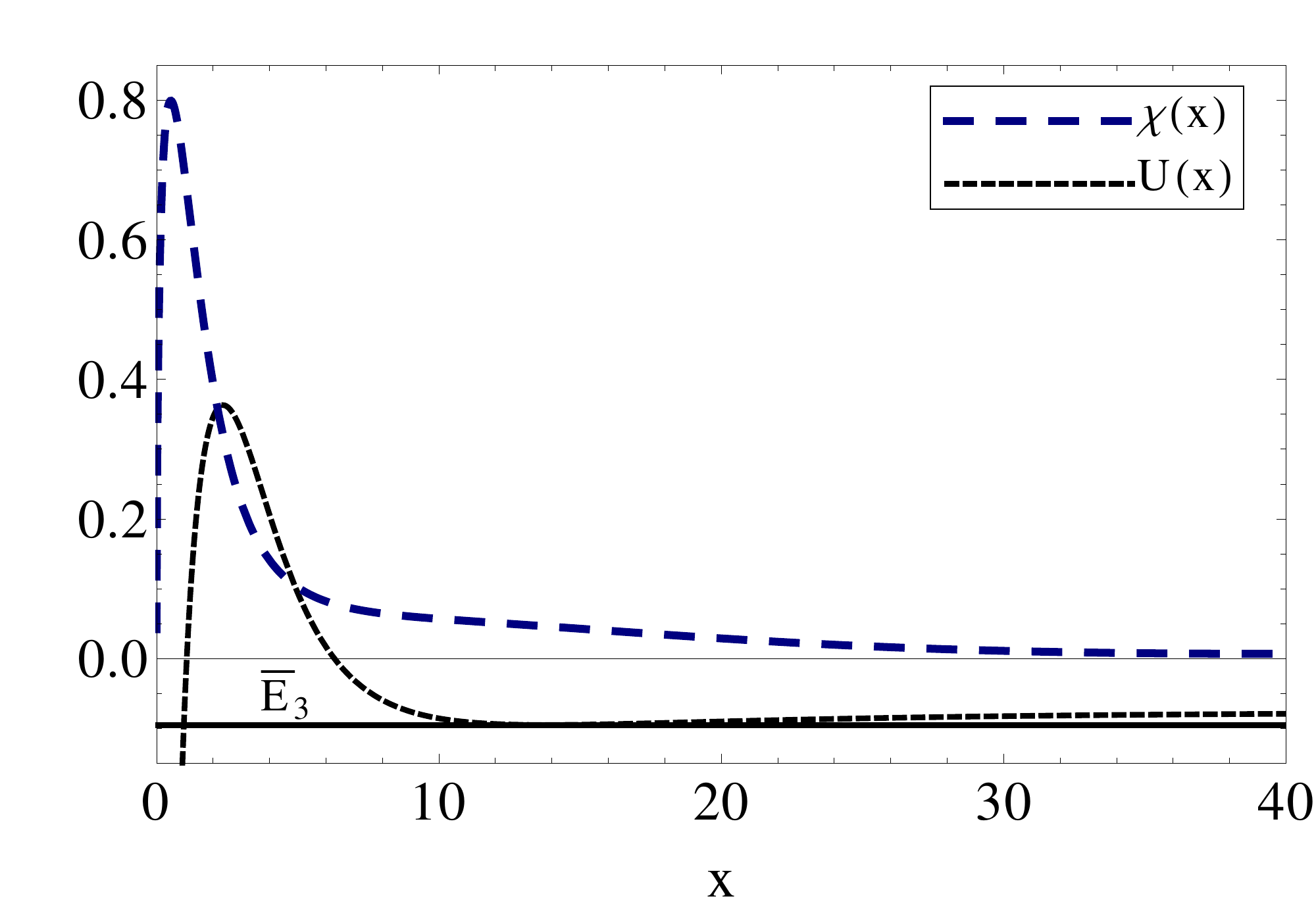}}\hspace{-0.25cm}
\subfloat[$\lambda  = 1.025$]{  \includegraphics[width=9cm]{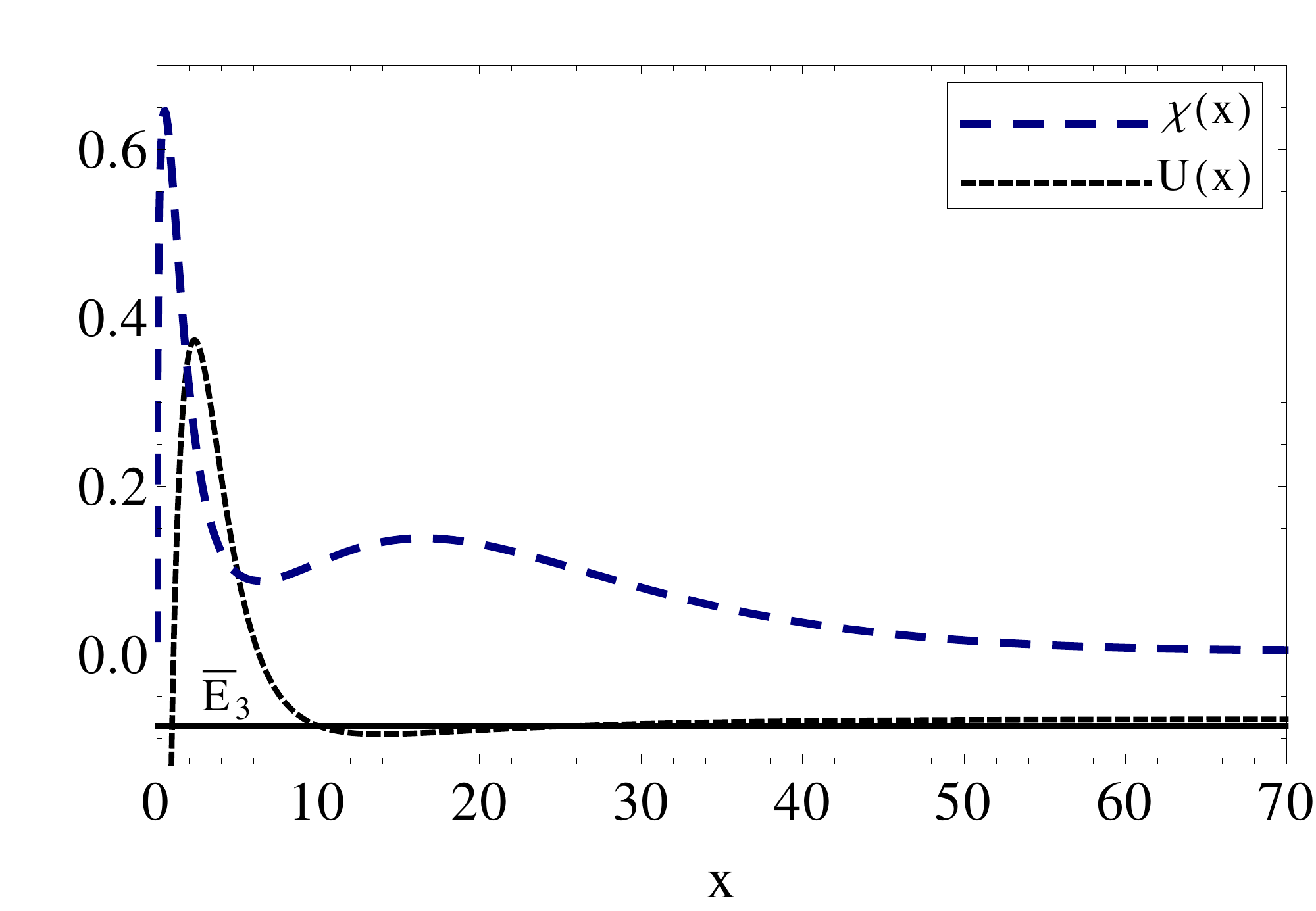}}
\\
\vspace{0.3cm}
\hspace{-0.8cm}\subfloat[$\lambda  = 1.035$]{  \includegraphics[width=9cm]{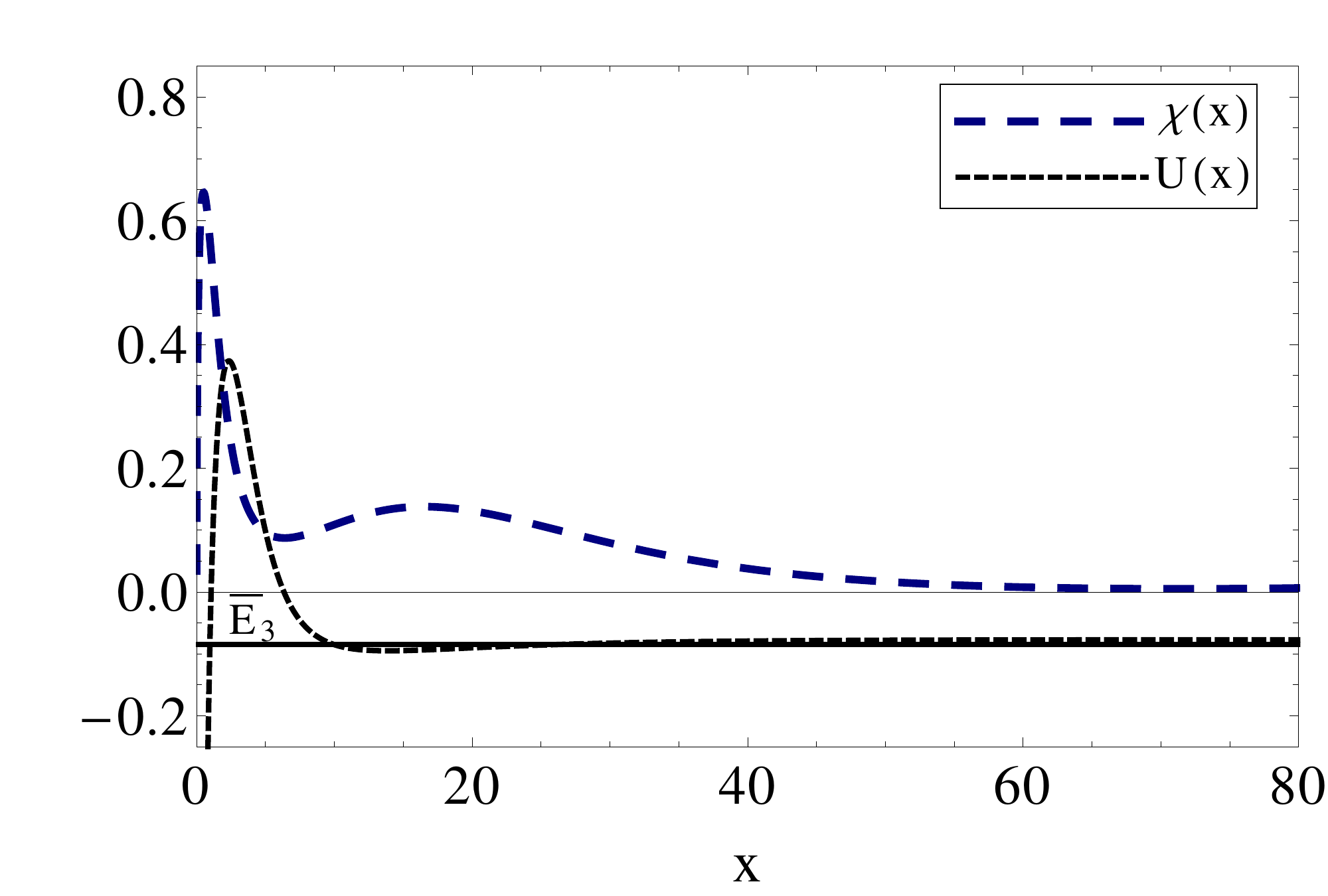}}\hspace{-0.25cm}
\subfloat[$\lambda  = 1.050$]{  \includegraphics[width=9cm]{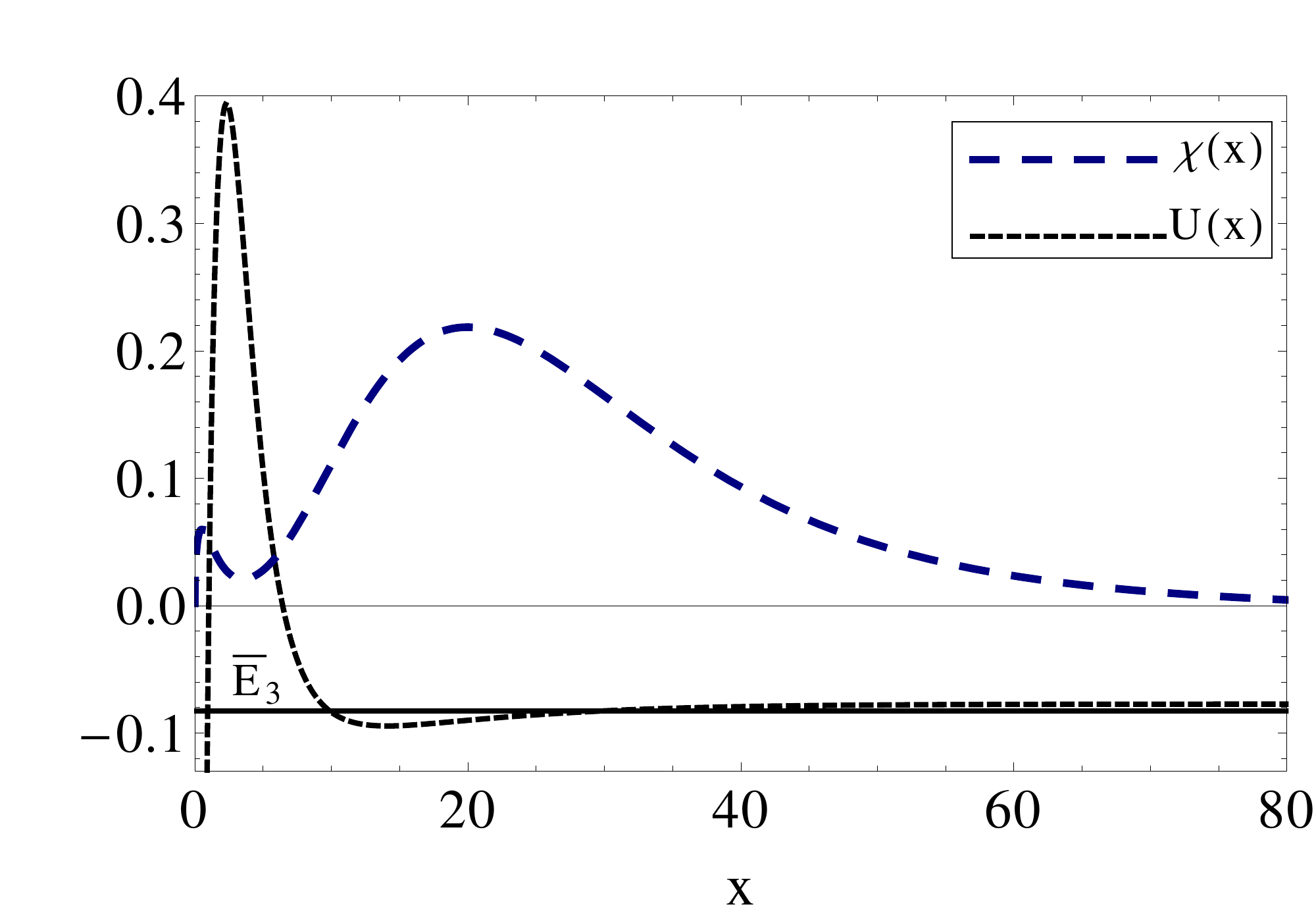}}
\caption{Sequence of figures showing how the total potential $U(x)$, 
 three-body energy ${\bar E}_3\equiv \left(a_0/2\mathcal{Q}_{\rm{eff}}^{2}\right)E_3$ and 
 the heavy-dipole ground-state wave function $\chi(x)$, change according to the strength of the dipole potential $\lambda$
 in the region where there is a more accentuated increase of $\langle R_{AA}^2\rangle$. The strength 
 is given just below of each figure, for a fixed $x_{0}=4.0$ and mass ratio $m_B / m_A =0.05$}
\label{figdiptot}
\end{figure*}

The relation between total potential, energy and wave function is
shown in Fig. \ref{figdiptot} for the ground state around the point of
avoided crossing.  The barrier of the potential is almost unchanged
with the height $ \approx 0.4 \left(2\mathcal{Q}_{\rm{eff}}^{2}/a_0\right)$
at $x \approx 2.2  $.  However, this only reflects that very small
variation of $\lambda$ results in a major change of the wave function.

For the smallest repulsion, the wave function is localized inside the
barrier, but quickly a tail develops and soon also a peak outside the
barrier.  Eventually this broad peak in the outer minimum carries all
the probability.  Tuning to the crossing point allows equal
probability in the two minima in this entangled state.  Very small
strength variation moves the probability to the inner or outer minima in
two corresponding clearly distinguishable states.  This size variation
is reflected in the mean-square radii, shown in Fig. \ref{figdip}.

At this point we want to emphasize that the dipole strength can be
simply changed by varying the direction of the dipole relative to the
two-dimensional plane where they move.  This possibility of
essentially manual control of the state, placing it on the left or
right of the barrier, simulates controlled entangled states, and may
thus provide a playground to simulate a qubit.

These properties depend on the chosen parameters in the calculations.
We have fixed the mass ratio, $m_{B}/m_{A}$, and the two-body strength
parameterized by $E_2$.  Decreasing $m_{B}/m_{A}$ or increasing $E_2$ must lead to
more attractive Born-Oppenheimer potentials: this in turn provides
more bound states for a given repulsive dipole strength, $\lambda$,
both in the inner but far more in the outer minimum.  However, avoided
crossings and the produced entangled states for very specific
$\lambda$-values would occur each time an inner and an outer localized
state become degenerate.  Thus, these properties occur
rather frequently and not necessarily among the lowest-lying levels.

\subsection{Critical dipole strength }

The total potential $U(x)$ as shown in Fig. \ref{potential} for
one dipole strength has a deep narrow minimum at small distances and a
very broad shallow minimum at much larger distances.  The details of
these characteristic features depend somewhat on the parameter
choices, that is mass ratio, $m_{B}/m_{A}$, two-body energy, $E_2$, and dipole strength
$\lambda$.  The number of bound states below the two-body continuum
threshold increases dramatically with decreasing $m_{B}/m_{A}$, but most of
these states would be localized in the outer minimum. 

The barrier created by the dipole repulsion would push the states
away, and the less room at small distance would severely limit the
corresponding number of bound states.  For any given mass ratio, the
increase in $\lambda$ pushes all energies towards the continuum
threshold.  First, all the small-distance states disappear and then
much more slowly one by one the large-distance localized states
also disappear.  For sufficiently large $\lambda$, no three-body bound
states are left even though the light particle is bound to each of the
heavy ones.  This happens for a critical strength value $\lambda = \lambda_c$,
in the dipole-dipole potential.

It is tempting to estimate $\lambda_c$ by use of the Landau criterion,
that is by having a negative two-dimensional volume of the potential.
We can integrate analytically both the Born-Oppenheimer and the dipole
potential.  The condition for vanishing volume below the two-body
threshold is then to integrate the total potential from
Eq. (\ref{totalpot}), equate to zero and determine $\lambda$; that is  
\begin{eqnarray} \label{landau}
 \int_0^{\infty} dx\ x \left[\frac{a_0}{2\mathcal{Q}_{\rm{eff}}^2}\bigg(\epsilon(x) 
 + |E_2|\bigg)  + \frac{\lambda x_0^3}{x^3+x_0^3} \right] = 0 \; .
\end{eqnarray}
The first of these integrals is obtained by  partial integration using the fact that $\lim_{x\rightarrow \infty} \epsilon(x) = -|E_{2}|$
\begin{eqnarray} \label{int}
& & \int_0^{\infty} dx\ x  \left[\frac{a_0}{2\mathcal{Q}_{\rm{eff}}^2}|E_{2}| (\bar{\epsilon}(x) 
 + 1) \right]\nonumber \\
 &=&  - \frac{a_0}{2\mathcal{Q}_{\rm{eff}}^2}|E_{2}| \int_0^{\infty} \frac{x^{2}}{2}\frac{d\bar{\epsilon}(x)}{dx}dx \;  ,
\end{eqnarray}
where $\bar{\epsilon}(x) \equiv \epsilon(x)/|E_{2}|$. Changing variable, $R = \left(a_{0}/2\right)x$ in Eq. (\ref{epsilon}) to write
\begin{equation}
 \ln \left[ \bar{\epsilon}(x)\right] = 2 K_{0}\left( y \right),
\end{equation}
where $y \equiv  \left(e^{\gamma} \mu_{B,AA}/2\mu_{AA}\right) \sqrt{\bar{\epsilon}(x)}\ x$ we can find an expression for $d\bar{\epsilon}(x) $ 
\begin{equation}
 d\bar{\epsilon}(x) = 2 \bar{\epsilon}(x) \frac{dK_{0}(y)}{dy}dy.
 \end{equation}
Finally, after a new partial integration, we write the contribution for the BO potential in the Landau criterion as
\begin{eqnarray}
&-4e^{-2\gamma} \frac{a_0}{2\mathcal{Q}_{\rm{eff}}^2}|E_{2}|  \left( \frac{\mu_{AA}}{\mu_{B,AA}}\right)\int_0^{\infty}dy\  y^{2}K^{\prime}_{0}(y) =  2.
\end{eqnarray}
The second integral in Eq. (\ref{landau}) gives
\begin{eqnarray}
 \int_0^{\infty} dx\ \frac{x}{x^3+x_0^3}  =   \frac{2\pi}{3\sqrt{3}x_0} \;.
\end{eqnarray}
Combining the results, we get the critical $\lambda$-value
\begin{eqnarray} \label{landau1}
 x_0^2 \lambda_c = \left(\frac{\mu_{AA}}{\mu_{B,AA}}\right) \frac{3\sqrt{3}}{\pi} \;.
\end{eqnarray}
Since $\mu_{AA} / \mu_{B,AA} \approx 0.5 m_A/m_B $ we get that $ x_0^2
\lambda_c \approx 0.83 m_A/m_B$, which predicts linear relation
between $\lambda_c$ and the mass ratio.  In Table \ref{critical} we
demonstrate that this prediction is far from being even qualitatively
correct.  On the other hand we find that the critical dipole strength
for constant $x_0$ accurately is quadratic in the mass ratio.

\begin{table}[h]
\caption{Critical strengths computed numerically $\lambda_{c}^{(N)}$, as
  function of mass ratio $m_B/m_A$ for $x_0 = 4.0$.}
\begin{tabular}{c c c}
\hline\hline
$m_{B}/m_{A}$ &\ \ \ \ \ \ \ \ \ $\left(\mu_{B,AA}/\mu_{AA}\right)\lambda_{c}^{(N)} $\ \ \ \ \ \ \ \ \ & $\left(\mu_{B,AA}\mu_{AA}\right)^{2}\lambda_{c}^{(N)}$   \\ [0.9ex]  
\hline 
0.001       & 17.891    & 0.0357 \\
0.005       &3.580  & 0.0357     \\
0.010        &1.790 & 0.0357     \\
0.050       &0.366  & 0.0357      \\
0.100        &0.191 & 0.0361     \\[1ex]
\hline\hline
\end{tabular}
\label{critical}  
\end{table}

The failure is easily explained since the potentials are far from being
weak as required for the Landau criterion to be valid.  The positive
contribution from the huge barrier has no influence on the outer
minimum where bound states can exist completely independent of the
small distance behavior.  This is related to the rather short range of
the dipole potential. Note that a short-distance potential like a Gaussian 
would make it even worse. The critical strength for such potentials can 
only be related to the sizes of their tails at distances where the bound 
states are localized.

Thus, the properties of the outer minimum are decisive.  We then
investigate the total potential for large $R$. It is convenient to use
the dimensionless measure, $s(R)$, defined in \cite{BelJPB2013}, that
is
\begin{eqnarray} \label{srrelat}
 s(R)  =   \sqrt{\frac{2\mu_{B,AA} |E_2|}{\hbar^{2}}}R   \;.
\end{eqnarray}
The total potential below $E_2$, $U(x)-\left(a_0/2\mathcal{Q}_{\rm{eff}}^{2} \right)|E_2|$, 
in dimensionless units, using the approximation for large values of 
$R$ (or $s(R)$ or $x(R))$ in the Born-Oppenheimer potential \cite{BelJPB2013}, is written as
\begin{eqnarray} \label{asymppot}
U(s)-\frac{a_0}{2\mathcal{Q}_{\rm{eff}}^{2}}|E_2|&=&\frac{a_0}{2\mathcal{Q}_{\rm{eff}}^{2}}|E_2|
\left[-1-\sqrt{2\pi}\frac{e^{-s}}{\sqrt{s}}\right]\nonumber \\
&+&\lambda x_0^3
\left(\frac{a_0}{2s} \sqrt{ \frac{ 2\mu_{B,AA}|E_2| } {\hbar^{2}} } \right)^3,\nonumber
\end{eqnarray}
\begin{eqnarray}
U(s)&=&\frac{e^{2\gamma}}{4}\left(\frac{\mu_{B,AA}}{\mu_{AA}}\right)
\left[  \left(\frac{\mu_{B,AA}}{\mu_{AA}}\right)^2
\frac{e^{\gamma}}{2 s^3}\ \lambda x_0^3\right. \nonumber \\
&-&\left. \sqrt{2\pi} \frac{e^{-s}}{\sqrt{s}} \right] \;,
\end{eqnarray}
where we assumed that $x \gg x_0$ and replaced the asymptotic form of $\epsilon(x)$ 
for large arguments.  Note that we replaced $\mathcal{Q}_{\rm{eff}}$ and $a_0$, 
respectively, by Eqs. (\ref{qeff}) and (\ref{aff}).  When both Eq. (\ref{asymppot}) 
and its derivative with respect to $s$ are zero at the same point in
space, the total potential has a minimum value which precisely is
above the infinitesimally small attraction sufficient to support a
bound state.  This potential is non-negative for all large values of
$s$.  The solutions to these two equations are $s=5/2$ and $\lambda =
\lambda_c$, where
\begin{eqnarray} \label{crit1}
  \lambda_c =  \left(\frac{\mu_{AA}}{\mu_{B,AA}}\right)^2 
 \frac{2\sqrt{2\pi}}{x_0^3e^{\gamma}} s^{5/2} e^{-s} \;.
\label{lc}
\end{eqnarray}
Inserting $s=5/2$ and the numerical values we then obtain
\begin{eqnarray} \label{crit2}
  \lambda_c = 0.035676\ \left(\frac{\mu_{AA}}{\mu_{B,AA}}\right)^2 \;,
\end{eqnarray}
which is precisely obtained by the explicite numerical calculation
shown in Table \ref{critical}.

The precision in this prediction is remarkable.  The analytic result
is obtained by assuming that even an infinitesimally small attraction
at large distance would be able to support a bound state.  It seems to
be violating the Landau criterion which is a global property for
globally weak potentials.  We interprete the above result by arguing
that the possibly large positive potentials at small distance are
irrelevant since that part cannot provide any binding. We can then as
well change the total potential to be zero in all space except at the
large distances where a small attractive pocket is left for
sufficiently large $\lambda$.  Certainly this new potential would be
globally very weak since it is zero except at large distance where it
is exponentially or cubically small.  Since the new potential also
must have more bound states than the initial potential, we can use the
original Landau argument and determine when there is not even an
infinitesimally small attraction left.

\subsection{Experimental possibilities}

In this subsection we compare our results with real systems 
to see whether they are experimentally feasible. 
The heavy subsystem, generically represented by the letter $A$, may 
be a magnetic or electric dipole. Similarly to the case of neutral 
atoms, where we can define a characteristic van der Waals length 
($\ell_{\rm vdW}$) that divides the fields of chemistry and 
physics \cite{BraPR2006}, we may also define a characteristic dipole 
length, $R_0$ \cite{BohNJP2009}. Both lengths, as well as the Bohr radius 
($a_0^B$), are defined equating the typical kinetic energy $\hbar^2/(\mu_{AA}L^2)$ 
and the potential energies: Coulomb $e^2/L$ ($L\equiv a_0^B$), van der Waals 
$C_6/L^6$ ($L\equiv \ell_{\rm vdW}$) and dipole $CD^2/L^3$ ($L\equiv R_0$). 
Thus, the dipole length is given by $R_0=\mu_{AA}CD^2/\hbar^2$.

The dipole length $R_{0}$ is very different for each system 
in such a way we cannot define an overall dipole length scale. However, using 
the dipole length for each system, we may study what the 
atom-dipole scattering length would be to unbind the three-body system using 
the critical $\lambda$ derived in the previous section.

From Eqs. (\ref{ux}) and (\ref{lc}) and writing explicitly the effective charge and 
radius given, respectively, by Eqs. (\ref{qeff}) and (\ref{aff}), we may write the 
ratio
\begin{eqnarray}
\left(\frac{\lambda}{\lambda_{c}} \right)&=& \frac{2}{\sqrt{\pi}s^{5/2}e^{-s}}
\left(\frac{|E_{2}|\mu_{B,AA}x_{0}^2}{\hbar^{2}}\right)^{3/2}\frac{CD^{2}}{|E_{2}|}\\
&=&\frac{64e^{-3\gamma}}{\sqrt{2\pi}s^{5/2}e^{-s}}
\left(\frac{|E_{2}|\mu_{AA}R_{0}}{\hbar^{2}}\right)^{3}\frac{CD^{2}}{|E_{2}|},
\end{eqnarray}
where $x_0=2R_0/a_0$ and the two-body binding energy $|E_2|$ in two dimensions for the 
shallow dimer in the unitary limit is given by \cite{BraPR2006} 
\begin{eqnarray}
|E_{2}| = 4e^{-2 \gamma}\frac{\hbar^{2}}{\mu_{AB}a^{2}},
\end{eqnarray}
where $a$ is the two-body scattering length in 2D. For a magnetic dipole the constants are 
given by $C = 1/c^{2}$ and $D^{2} = \mu^{2}$, where the dipole moment $\mu$ should be 
in units of the Bohr magneton, $\mu_{B}{\equiv}(e\hbar)/(2m_{e})$. Then, 
the ratio is given by 
\begin{eqnarray}
\nonumber
&&\left(\frac{\lambda}{\lambda_c}\right)_{{\rm MAG}}=\frac{256e^{(s-7\gamma)}\mu^2\mu_{AA}^3R_0^3}
{137^2\sqrt{2\pi}s^{5/2}\mu_{B}^{2}\mu_{AB}^2m_ea_0^{B3}}\left(\frac{a_0^B}{a}\right)^4,
\end{eqnarray}
where $s=5/2$.

For an electric dipole, we have $C = 1$ and $D^{2} = d^{2}$, where the dipole moment $d$ should be in 
atomic units $d_{0} \equiv e a_{0}^{B}$. Then, the ratio for an electric dipole 
reads:
\begin{eqnarray}
\nonumber
&&\left(\frac{\lambda}{\lambda_c}\right)_{{\rm EL}}=\frac{1024e^{(s-7\gamma)}d^2\mu_{AA}^3R_0^3}
{\sqrt{2\pi}s^{5/2}d_{0}^{2}\mu_{AB}^2m_ea_0^{B3}}\left(\frac{a_0^B}{a}\right)^4.
\end{eqnarray}

These $\lambda$-ratios increase with dipole moment, length and mass,
while decreasing with light mass and heavy-light scattering length.
Thus, for a given three-body system we can determine which scattering
length, $a$, must be exceeded to allow any bound state structure.
The tunneling effect occurs for a range of smaller $\lambda$'s.
It is worth mentioning that the tunneling effect, and 
consequently the avoid crossings, does not happen for all values of $x_0$ used to regularize 
the dipole potential at short distances. In order to have the tunneling, $x_0$ should roughly 
be smaller than 7 as for larger values the minimum located at large distances 
disappears and there is no bound state on the right side of the barrier. $x_0$ is given by 
\begin{eqnarray}
x_{0} = 4\sqrt{2}e^{- 2\gamma} \left(\frac{R_{0}}{a}\right)\frac{\mu_{AA}}{\sqrt{\mu_{B,AA} \mu_{AB}}}.
\label{x0}
\end{eqnarray}

Let us consider some systems where the light particle is ${^6}$Li and the
two identical heavy particles are either molecules with electric
dipoles or atoms with magnetic dipoles. These systems are listed in 
Table \ref{tableexp} with respective dipole moments, lengths and masses 
along with the limiting two-body scattering lengths to bind the three-body systems 
($\lambda=\lambda_c$) and the $x_0$-values corresponding to the short-range parameter 
that regularizes the dipole potential.

We conclude that although some of these scattering lengths are rather
large they are all within present-day experimental possibilities and,
furthermore, the $x_0$-values are all inside the range for tunneling 
to be observed.

\begin{table}[h!]
\centering
\caption{Experimental properties of each dipole $A$ represented in the first column. 
The second column shows the values of the electric or magnetic dipole moments. 
Electric dipole moments are given in Debye (D) units, 1 Debye$=0.3934ea_0^B$ 
($e$ is the electron charge, $a_0^B=\hbar^2/(m_ee^2)=0.53$ \AA\, 
is the Bohr radius and $m_e$ the mass of the electron), and magnetic dipole 
moments are given in units of $\mu_B$. In the third column we 
give the dipole length in units of $a_0^B$. The last two columns give the scattering 
length that gives $\lambda=\lambda_c$ and the value of $x_0$ used to regulate 
the dipole potential at short distances.}
\begin{tabular}{l c c c c}
\hline\hline
{\rm Dipole(A)}                  			  & Moments      & $R_{0}/a_{0}^{B}$ 		  &\ \ \ \  $a/a_{0}^{B}$        	    & $x_{0}$       \\[0.9ex]  
\hline  
Singlet $^{40}$K-$^{87}$Rb \cite{NiScience2008}  	  & 0.57 D &    $4	\times10^{3}$	        &              	25275.5		                &     3.12    \\
$^{87}$Rb-$^{133}$Cs   \cite{DanThesis} 		 & 1.25 D	&	$3	\times10^{4}$	&              253717		  	       &       3.97   	 \\
$^{87}$Rb    \cite{DanThesis}				 & 0.5 $\mu_{B}$		&	0.2			&              1.0		  	       &       2.7   	 \\
$^{133}$Cs \cite{DanThesis}				 &   0.75 $\mu_{B}$	&	0.6			 &             	3.9       			&      3.16    	\\
$^{164}$Dy \cite{TanPRA2015}   \ \      	 &  10 $\mu_{B}$ 	 &	195			&        	1275				 &       3.86  	\\       
$^{52}$Cr   \cite{GriPRL2006} \ \   	& 6 $\mu_{B}$		&       15			 &             63.23	  			  &       2.05    	     \\
$^{167}$Er$_{2}$  \cite{FriPRL2015}  \ \  		&  14 $\mu_{B}$		&	1600			 &              12357.1				 &      6.5        \\[1ex]
\hline\hline
\end{tabular}
\label{tableexp}  
\end{table}

\section{Summary and Conclusions}
\label{conclusion}

We studied a three-body system composed of one light and two heavy
dipoles, all confined to two spatial dimensions.  We used a
zero-range interaction between the light and each of the two identical heavy
dipoles.  This extremely short-range interaction has several
advantages: first, it serves as a schematic prototype of a
short-range interaction; second, the properties can be semi-analytically
derived; and third, the results are universal by definition.  The strength is
parametrized in terms of the light-heavy system binding energy which
then is one of our input parameters.  We limit ourselves to a small
light-heavy mass ratio allowing the use of the Born-Oppenheimer
approximation for an effective potential between the two heavy
particles.

The two slowly moving heavy dipoles may also interact directly in
addition to the Born-Oppenheimer potential.  Here we only considered
the realistic possibility of the dipole-dipole interaction which has
an inverse cubic dependence on their separation.  The strength of this
interaction is, besides the size of the dipole moment, determined by
the direction of the dipole moments relative to the two-dimensional
planar confinement. Adjusting this orientation, which can be achieved 
experimentally through external fields, the interaction can vary from 
an attractive to a repulsive potential passing through zero. We have in 
this work essentially considered only the repulsive and vanishing potentials.

After having established the formalism, we discussed the three-body
properties emerging from the Born-Oppenheimer potential and vanishing
direct heavy-heavy interaction.  The light particle generates an
effective interaction as a consequence of its exchange between the two
heavy particles.  Therefore, the lighter the atom the easier it can be
exchanged which then increases the heavy-heavy effective attraction.
The number of excited states then increases with decreasing mass
ratio, and it is in this way controlled by the constituents of the trap.

Through the Born-Oppenheimer procedure we derived an analytic
expression for the wave function depending on the three-body Jacobi
relative coordinates. We rewrite and exhibit the wave function in
terms of two dimensionless coordinates and a dimensionless scale
parameter.  The logarithmic divergence is clearly seen when one of the
heavy particles is on top of the light one.  This coordinate-space
wave function may be used in other ``realistic'' calculations as input
to parameterize the large-distance universal tail of the wave function.

The effective potential is Coulomb-like at small distances with an effective
charge proportional to the square root of the two-body energy divided by
the light particle mass.  The corresponding Coulomb degeneracy arising
from the angular momentum and radial node quantum numbers is very accurate
for the lowest-lying well bound states.  The systematic deviation from
the Coulomb results is traced to the next order attractive correction
terms which are a sum of a constant and a slowly varying logarithmic
term.  The Coulomb-like behavior is slowly destroyed as the energies 
approach the two-body continuum threshold.

We included the dipole-dipole interaction, oriented to produce
repulsion, and minimally regularized for zero separation.  The total
heavy-heavy potential is now the result of a competition between the
two contributions.  At the smallest distances, the Coulomb-like
attraction survives with a translated energy scale. For moderate
repulsive strengths, a barrier appears before a broad minimum at
larger distances.  Increasing the repulsion affects the short-distance
localized states much more than states in the broad minimum. Specific
strengths therefore lead to crossing of the inner and outer localized
states. At the avoided crossing points two levels are nearly degenerate
and each is a mixture of the two structures.  Characteristic signals of this
type of tunneling phenomenon are the avoided crossing spectrum, and an
exchange of properties such as mean square radii between the two levels
involved.

Increasing the heavy-heavy repulsion sufficiently the bound states disappear one
by one until no bound state is left and the three-body system is
unstable towards emission of one of the heavy particles.  This defines
a critical dipole strength as a function of mass ratio and two-body
strength.  We derived the Landau criterion analytically for this
critical strength, but concluded that it is wrong both quantitatively
as well as qualitatively.  The relatively short-range nature of the
dipole repulsion requires a large strength to exclude all bound
states.  The resulting large positive barrier at intermediate
distances violates the assumption of a weak potential for validity of
the Landau criterion.  The last bound state owes its existence to a
tiny attraction at the outer edge of the exponentially decaying
Born-Oppenheimer potential.  When this pocket is wiped out, the
repulsive dipole has just passed the critical strength, which we
analytically and numerically calculate to be proportional to the
square of the inverse mass ratio.

The full Faddeev calculation \cite{SaPRA2016} is numerically feasible
in 2D.  However, the Born-Oppenheimer approximation separates fast and
slow motion and provides semi-analytic results which in simple terms
explain the underlying complicated physics.  This is especially
highlighted in the Coulomb-like short-distance part of the effective
potential which explicitly shows the physical reason for the increase
of bound states when the mass ratio is decreased.  This also shows the
crucial difference from the $1/r^2$ potential in 3D which is
responsible for the appearance of the scaling laws characteristic of
Efimov states.

The Born-Oppenheimer established attractive $1/r$ effective
short-range potential reveals the difference between studies using the
present dipolar potential and zero-range interactions
\cite{BelJPB2013,SaPRA2016}.  Tuning the strengths of the finite-range
(dipolar) and the short-range Born-Oppenheimer potentials can
emphasize short or long-range behavior, or allowing equal influence.
This opens a field of new physics which depends on both the strength and
shape of the finite-range interaction.  This variation generates
interesting phenomena in energy spectra and mean-square radii that is
not possible when assuming only attractive short-range interactions.
Consequently, the results are no longer universal.

The study made in this paper can be tested experimentally and easily
extended to three distinguishable particles.  The possibility to
localize degenerate wave functions in the two minima by tuning the
interactions may provide an interesting alternative to controlled
transfer of quantum information between these entangled states.  
The possibility to vary the strength by changing
angles of the dipoles combined with the Feshbach resonance technique
to control the short-range interaction produces a very rich playground
to study few-body correlations.

\acknowledgments
This work was partly supported by funds provided by the Brazilian agencies 
Funda\c{c}\~{a}o de Amparo \`{a} Pesquisa do Estado de S\~{a}o Paulo - FAPESP grants no. 
2016/01816-2(MTY) and 2013/01907-0(GK), Conselho Nacional de Desenvolvimento 
Cient\'{i}fico e Tecnol\'{o}gico - CNPq grant no. 305894/2009(GK) and 302701/2013-3(MTY), 
Coordena\c{c}\~{a}o de Aperfei\c{c}oamento de Pessoal de N\'{i}vel Superior - 
CAPES no. 88881.030363/2013-01(MTY) and 33015015001P7(DSR).

\end{document}